\documentclass[twocolumn,amsmath,amssymb]{revtex4}

\usepackage{amsmath}
\usepackage{graphicx}

\usepackage{amssymb}
\usepackage{amsmath}

\begin{document}



\title{Strain-engineering of graphene's electronic structure beyond continuum elasticity}

\author{Salvador Barraza-Lopez}
\address{Department of Physics. University of Arkansas. Fayetteville, AR 72701, USA}
\email{sbarraza@uark.edu}
\author{Alejandro A. Pacheco Sanjuan}
\address{Departamento de Ingenier{\'\i}a Mec\'anica. Universidad del Norte. Km.~5 V{\'\i}a Puerto Colombia. Barranquilla, Colombia}
\author{Zhengfei Wang}
\address{Department of Materials Science and Engineering. University of Utah. Salt Lake City, UT 84112, USA}
\author{Mihajlo Vanevi\'c}
\address{Department of Physics, University of Belgrade. Studentski trg 12, 11158 Belgrade, Serbia}

\begin{abstract}
We present a new first-order approach to strain-engineering of graphene's electronic structure where no continuous displacement field $\mathbf{u}(x,y)$ is required. The approach is valid for negligible curvature. The theory is directly expressed in terms of atomic displacements under mechanical load, such that one can determine if mechanical strain is varying smoothly at each unit cell, and the extent to which sublattice symmetry holds. Since strain deforms lattice vectors at each unit cell, orthogonality between lattice and reciprocal lattice vectors leads to renormalization of the reciprocal lattice vectors as well, making the $K$ and $K'$ points shift in opposite directions. From this observation we conclude that no $K-$dependent gauges enter on a first-order theory. In this formulation of the theory the deformation potential and pseudo-magnetic field take discrete values at each graphene unit cell. We illustrate the formalism by providing strain-generated fields and local density of electronic states on graphene membranes with large numbers of atoms. The present method complements and goes beyond the prevalent approach, where strain engineering in graphene is based upon first-order continuum elasticity.
\end{abstract}
\date{Available online: 14 May 2013}
\pacs{A. graphene membranes \sep C. Electronic structure \sep D. Elasticity theory}
\maketitle


\section{Introduction}

The interplay between mechanical and electronic effects in carbon nanostructures has been studied for a long time (e.g., \cite{Ando2002,GuineaNatPhys2010,castroRMP,Pereira1,Vozmediano,deJuanPRL2012,Asgari,r2,Peeters1,Peeters2,Peeters3}). The mechanics in those studies invariably enters within the context of continuum elasticity. One of the most interesting predictions of the theory is the creation of large, and roughly uniform pseudo-magnetic fields and deformation potentials under strain conformations having a three-fold symmetry \cite{GuineaNatPhys2010}. Those theoretical predictions have been successfully verified experimentally \cite{Crommie,Gomes2012}.

Nevertheless, different theoretical approaches to strain engineering in graphene possess subtle points and apparent discrepancies \cite{deJuanPRL2012,Kitt2012}, which may hinder progress in the field. This motivated us to develop an approach \cite{us} which does not suffer from limitations inherent to continuum elasticity. This new formulation accommodates numerical verifications to determine when arbitrary mechanical deformations preserve sublattice symmetry. Contrary to the conclusions of Ref.~\cite{Kitt2012}, with this formulation one can also demonstrate in an explicit manner the absence of $K-$point dependent gauge fields on a first-order theory (see Refs.~\cite{us} and \cite{arxiv,Kitt2} as well). The formalism takes as its only direct input {\em raw} atomistic data --as the data obtained from molecular dynamics runs. The goal of this paper is to present the method, making the derivation manifest. We illustrate the formalism by computing the gauge fields and the density of states in a graphene membrane under central load.

\begin{figure}[tb]
\includegraphics[width=0.45\textwidth]{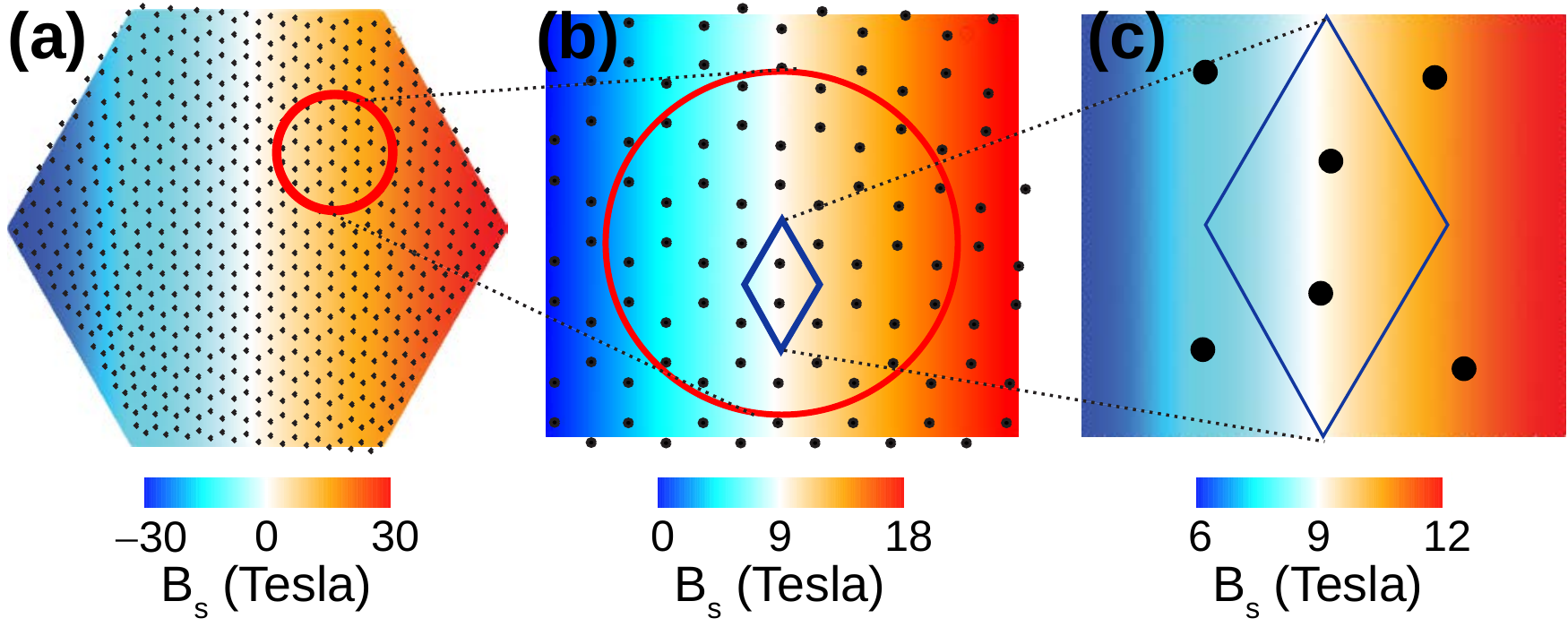}
\caption{Gauge fields from first-order continuum elasticity are defined regardless of spatial scale. A unit cell is shown in (b) and (c) for comparison. In this work, we define the pseudospin Hamiltonian for each unit cell using space-modulated, low-energy expansions of a tight-binding Hamiltonian in reciprocal space. As a result, in our approach the gauge fields will become discrete.}\label{fig:F1}
\end{figure}

\subsection{Motivation}
 The theory of strain-engineered electronic effects in graphene is semi-classical. One seeks to determine the effects of mechanical strain across a graphene membrane in terms of spatially-modulated pseudospin Hamiltonians $\mathcal{H}_{ps}$; these pseudospin Hamiltonians $\mathcal{H}_{ps}(\mathbf{q})$  are low-energy expansions of a Hamiltonian formally defined in reciprocal space. Under ``long range'' mechanical strain (extending over many unit cells and preserving sublattice symmetry \cite{Ando2002,GuineaNatPhys2010,castroRMP})  $\mathcal{H}_{ps}$ also become continuous and slowly-varying local functions of strain-derived gauges, so that $\mathcal{H}_{ps}\to\mathcal{H}_{ps}(\mathbf{q},\mathbf{r})$. Within this first-order approach, the salient effect of strain is a local shift of the $K$ and $K'$ points in opposite directions, similar to a shift induced by a magnetic field \cite{GuineaNatPhys2010,castroRMP}. In the usual formulation of the theory \cite{Ando2002,GuineaNatPhys2010,castroRMP,Pereira1,Vozmediano,deJuanPRL2012}, this dependency on position leads to a {\em continuous} dependence of strain-induced fields $\mathbf{B}_s(\mathbf{r})$ and $E_s(\mathbf{r})$. Such continuous fields are customarily superimposed to a discrete lattice, as in Figure \ref{fig:F1}~\cite{GuineasSSC2012}.

 When expressed in terms of continuous functions, a pseudospin Hamiltonian $\mathcal{H}_{ps}$ is defined down to arbitrarily small spatial scales and it spans a zero area. In reality, however, the pseudospin Hamiltonian can only be defined per unit cell, so it should take a single value at an area of order $\sim a_0^2$  ($a_0$ is the lattice constant in the absence of strain).

This observation tells us already that the scale of the mechanical deformation with respect to a given unit cell is inherently lost in a description based on a continuum model. For this reason, it is important to develop an approach which is directly related to the atomic lattice, as opposed to its idealization as a continuum medium. In the present paper we show that in following this program one gains a deeper understanding of the interrelation between the mechanics and the electronic structure of graphene. Indeed, within this approach we are able to quantitatively analyze whether the proper phase conjugation of the pseudospin Hamiltonian holds at each unit cell. The approach presented here will give (for the first time) the possibility to explicitly check on any given graphene membrane under arbitrary strain if mechanical strain varies smoothly on the scale of interatomic distances. Consistency in the present formalism will also lead to the conclusion that in such scenario strain will not break the sublattice symmetry but the Dirac cones at the $K$ and $K'$ points will be shifted in the opposite directions \cite{GuineaNatPhys2010,castroRMP}.

 Clearly, for a reciprocal space to exist one has to preserve crystal symmetry, so that when crystal symmetry is strongly perturbed, the reciprocal space representation starts to lack physical meaning, presenting a limitation to the semiclassical theory. The lack of sublattice symmetry --observed on actual unit cells on this formulation beyond first-order continuum elasticity-- may not allow proper phase conjugation of pseudospin Hamiltonians at unit cells undergoing very large mechanical deformations. Nevertheless this check cannot proceed --and hence has never been discussed-- on a description of the theory within a continuum media, because by construction there is no direct reference to actual atoms on a continuum.

 As it is well-known, it is also possible to determine the electronic properties directly from a tight-binding Hamiltonian $\mathcal{H}$ in real space, without resorting to the semiclassical approximation and without imposing an {\em a priori} sublattice symmetry. That is, while the semiclassical $\mathcal{H}_{ps}(\mathbf{q},\mathbf{r})$ is defined in reciprocal space (thus assuming some reasonable preservation of crystalline order), the tight-binding Hamiltonian $\mathcal{H}$ in real space is more general and can be used for membranes with arbitrary spatial distribution and magnitude of the strain.

 In addition, contrary to the claim of Ref.~\cite{Kitt2012}, the purported existence of $K-$point dependent gauge fields does not hold on a first-order formalism \cite{us,arxiv}. What we find instead, is a shift in opposite directions of the $K$ and $K'$ points upon strain~\cite{GuineaNatPhys2010}.

\section{Theory}
\subsection{Sublattice symmetry}

The continuum theories of strain engineering in graphene, being semiclassical in nature, require sublattice symmetry to hold \cite{Ando2002,GuineaNatPhys2010}.
One the other hand, no measure exists in the continuum theories \cite{Ando2002,GuineaNatPhys2010,castroRMP,Pereira1,Vozmediano,deJuanPRL2012} to test sublattice symmetry on actual unit cells under a mechanical deformation. For this reason, sublattice symmetry is an implicit assumption embedded in the continuum approach.

\begin{figure}[h!]
\includegraphics[width=0.45\textwidth]{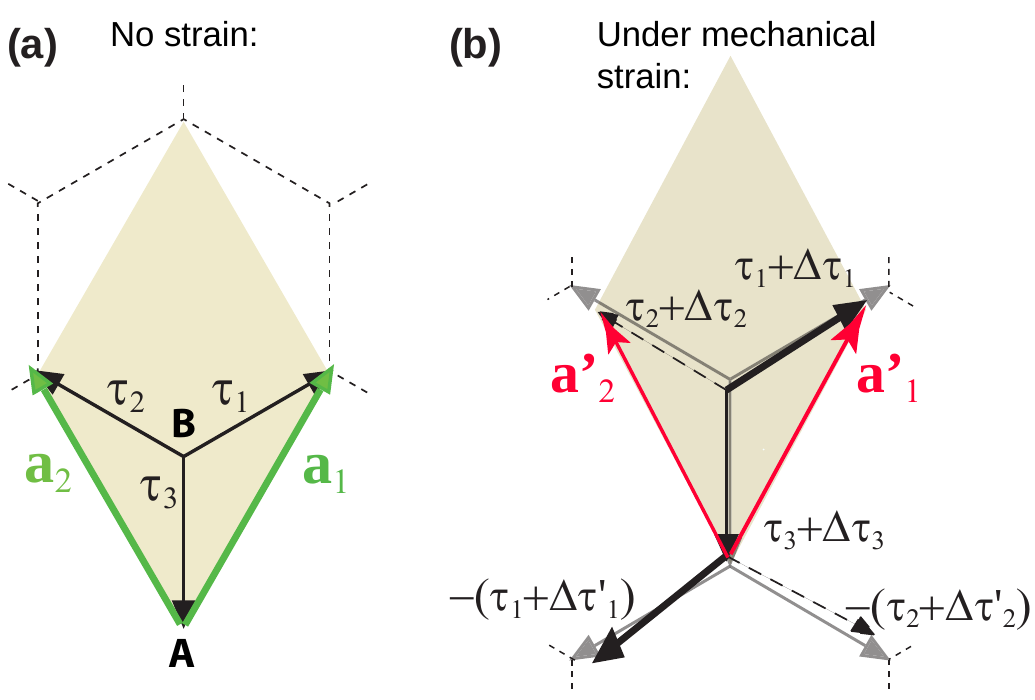}
\caption{(a) Definitions of geometrical parameters in a unit cell. (b) Sublattice symmetry relates to how {\em pairs} of nearest-neighbor vectors (either in thick, or dashed lines) are modified due to strain. These vectors change by $\Delta \mathbf{\tau}_j$ and $\Delta \mathbf{\tau}_j'$ upon strain ($j=1,2$). Relative displacements of neighboring atoms lead to modified lattice vectors; the choice of renormalized lattice vectors will be unique {\em only} to the extent to which sublattice symmetry is preserved: $\Delta \mathbf{\tau}_j'\simeq \Delta \mathbf{\tau}_j$.}\label{fig:F2}
\end{figure}

To address the problem beyond the continuum approach, let us start by considering the unit cell before (Fig.~\ref{fig:F2}(a)) and  after arbitrary strain has been applied (Fig.~\ref{fig:F2}(b)). For easy comparison of our results, we make the zigzag direction parallel to the $x-$axis, which is the choice made in Refs.~\cite{GuineaNatPhys2010} and \cite{Vozmediano}. (Arbitrary choices of relative orientation are clearly possible; in Ref.~\cite{us} we chose the zigzag direction to be parallel to the y-axis.)

The lattice vectors before the deformation are given by (Fig.~\ref{fig:F2}(a)):
 \begin{equation}\label{eq:defa}
 \mathbf{a}_1=\left(1/2,\sqrt{3}/2\right)a_0,\text{ }\mathbf{a}_2=\left(-{1}/{2},{\sqrt{3}}/{2}\right)a_0,
 \end{equation}
 \begin{equation}\label{eq:deft}
 \boldsymbol{\tau}_1=\left(\frac{\sqrt{3}}{2},\frac{1}{2}\right)\frac{a_0}{\sqrt{3}},\text{ } \boldsymbol{\tau}_2=\left(-\frac{\sqrt{3}}{2},\frac{1}{2}\right)\frac{a_0}{\sqrt{3}},\text{ }
 \boldsymbol{\tau}_3=\left(0,-1\right)\frac{a_0}{\sqrt{3}}.
 \end{equation}
 (Note that $\mathbf{a}_1=\boldsymbol{\tau}_1-\boldsymbol{\tau}_3$, and
 $\mathbf{a}_2=\boldsymbol{\tau}_2-\boldsymbol{\tau}_3$.)

After mechanical strain is applied (Fig.~\ref{fig:F2}(b)), each local pseudospin Hamiltonian will only have physical meaning at the unit cells where:
\begin{equation}\label{eq:applicabilitycondition}
\Delta \boldsymbol{\tau}_j'\simeq\Delta \boldsymbol{\tau}_j \text{ (j=1,2)}.
\end{equation}
 Condition (\ref{eq:applicabilitycondition}) can be re-expressed in terms of changes of angles $\Delta \alpha_j$ or lengths $\Delta L_j$ for pairs of nearest-neighbor vectors $\boldsymbol{\tau}_j$ and $\boldsymbol{\tau}_j'$
 [$j=1$ is shown in thick solid and $j=2$ in thin dashed lines in Fig.~\ref{fig:F2}(b)]:
\begin{equation}\label{eq:beta}
\small(\boldsymbol{\tau}_j+\Delta  \boldsymbol{\tau}_j)\cdot(\boldsymbol{\tau}_j+\Delta\boldsymbol{\tau}'_j)=
|\boldsymbol{\tau}_j+\Delta\boldsymbol{\tau}_j||\boldsymbol{\tau}_j+\Delta\boldsymbol{\tau}'_j|\cos(\Delta\alpha_j),
\end{equation}
\begin{equation}\label{eq:sign}
\small\text{sgn}(\Delta \alpha_j)=\text{sgn}\left([(\boldsymbol{\tau}_j+\Delta\boldsymbol{\tau}_j)
\times(\boldsymbol{\tau}_j+\Delta\boldsymbol{\tau}'_j)]\cdot \hat{k}\right),\end{equation}
where $\hat{k}$ is a unit vector along the z-axis, $sgn$ is the sign function ($sgn(a)=+1$ if $a\ge 0$ and $sgn(a)=-1$ if $a <0$), and:
\begin{equation}\label{eq:L}
\small
\Delta L_j\equiv |\boldsymbol{\tau}_j+\Delta\boldsymbol{\tau}_j|-|\boldsymbol{\tau}_j+\Delta\boldsymbol{\tau}'_j|.
\end{equation}

Even though in the problems of practical interest the deviations from the sublattice symmetry do tend to be small \cite{us}, it is important to bear in mind that the sublattice symmetry {\it does not hold a priori} \cite{GuineaNatPhys2010}. It is therefore important to have a method to quantify such deviations and check whether the sublattice symmetry holds at the problem at hand. Forcing the sublattice symmetry to hold from the start amounts to introducing an artificial mechanical constraint on the membrane which is not justified on physical grounds~\cite{Ericksen}. For this reason the method we propose is discrete and directly related to the actual lattice; it does not resort to the approximation of the membrane as a continuum medium \cite{Ando2002,GuineaNatPhys2010,castroRMP,Pereira1,Vozmediano,deJuanPRL2012,arxiv,Kitt2}. Being expressed in terms of the actual atomic displacements, our formalism holds beyond the linear elastic regime where the first-order continuum elasticity may fail. The continuum formalism is recovered as a special case of the one presented here in the limit when $|\Delta\mathbf{\tau}_j|/a_0\to 0$.

\subsection{Renormalization of the lattice and reciprocal lattice vectors}\label{sec:3}

\begin{figure}[h!]
\includegraphics[width=0.45\textwidth]{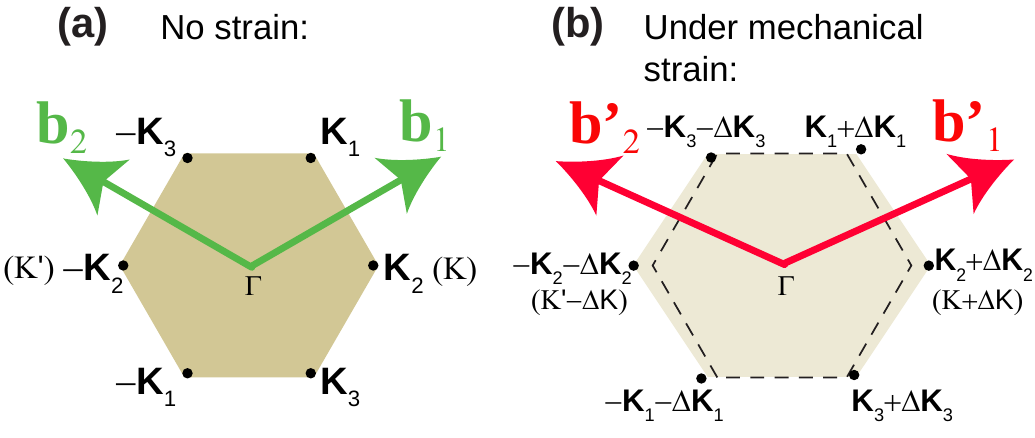}
\caption{First Brillouin zone (a) before and (b) after mechanical strain is applied. The reciprocal lattice vectors are shown,
 as well as the changes of  the high-symmetry points at the corners of the Brillouin zone. Note that independent $K$ points ($K$ and $K'$) move in the opposite directions. The dashed hexagon in (b) represents the boundary of the first Brillouin zone in the absence of strain.}\label{fig:F3}
\end{figure}

 In the absence of mechanical strain, the reciprocal lattice vectors $\mathbf{b}_1$ and $\mathbf{b}_2$ are obtained by standard methods: We define $\mathcal{A}\equiv(\mathbf{a}_1^T,\mathbf{a}_2^T)$, with $\mathbf{a}_1$ and $\mathbf{a}_2$ given in Eq.~(\ref{eq:defa}) and shown in Fig.~\ref{fig:F2}(a). The reciprocal lattice vectors $\mathcal{B}\equiv(\mathbf{b}_1^T,\mathbf{b}_2^T)$ are related to the lattice vectors by \cite{MartinBook}:
 \begin{equation}\label{eq:realreciprocal}
 \mathcal{B}^T=2\pi\mathcal{A}^{-1}.
 \end{equation}
 With the choice we made for $\mathbf{a}_1$ and $\mathbf{a}_2$ we get:
 \begin{equation}
 \mathbf{b}_1=\left(1,\frac{1}{\sqrt{3}}\right)\frac{2\pi}{a_0} \text{, and }
 \mathbf{b}_2=\left(-1,\frac{1}{\sqrt{3}}\right)\frac{2\pi}{a_0}.
 \end{equation}
  As seen in Fig.~\ref{fig:F3}(a) the $K-$points on the first Brillouin zone are defined by:
\begin{equation}
\mathbf{K}_1=\frac{2\mathbf{b}_1+\mathbf{b}_2}{3}, \text{ }\mathbf{K}_2=\frac{\mathbf{b}_1-\mathbf{b}_2}{3} \text{, and } \mathbf{K}_3=-\frac{\mathbf{b}_1+2\mathbf{b}_2}{3},
\end{equation}
and:
\begin{equation}
\mathbf{K}_4=-\mathbf{K}_1,\text{ } \mathbf{K}_5=-\mathbf{K}_2, \text{ and }\mathbf{K}_6=-\mathbf{K}_3.
\end{equation}

The relative positions between atoms change when strain is applied: $\boldsymbol{\tau}_j\to \boldsymbol{\tau}_j+\Delta\boldsymbol{\tau}_j$ ($j=1,2,3)$, and $-\boldsymbol{\tau}_j\to -\boldsymbol{\tau}_j-\Delta\boldsymbol{\tau}_j'$ ($j=1,2$).
For negligible curvature, one may assume that $\Delta\boldsymbol{\tau}_j\cdot\hat{z}=\Delta z_j\sim 0$ (and similar for the primed displacements $\Delta \boldsymbol{\tau}_j'$). We present here a formulation of the theory strictly valid for in-plane strain (it would also be valid for membranes with negligible curvature).

We wish to find out how reciprocal lattice vectors change to first order in displacements under mechanical load.
 In order for reciprocal lattice vectors to make sense at each unit cell, Eqn.~\ref{eq:applicabilitycondition} must hold. In terms of numerical quantities one would need that $\Delta \alpha_j$ and $\Delta L_j$ are all close to zero. In that case we set $\Delta \boldsymbol{\tau}_j'\to \Delta \boldsymbol{\tau}_j$ for j=1,2, and continue our program.

 For this purpose we define:
\begin{equation}
\Delta \mathbf{a}_1\equiv\Delta \boldsymbol{\tau}_1-\Delta \boldsymbol{\tau}_3 \text{, and }
\Delta \mathbf{a}_2\equiv\Delta \boldsymbol{\tau}_2-\Delta \boldsymbol{\tau}_3,
\end{equation}
or in terms of (two-dimensional) components:
\begin{equation}
\Delta \mathcal{A}\equiv
\left(
\begin{matrix}
\Delta \tau_{1x}-\Delta \tau_{3x}& \Delta \tau_{2x}-\Delta \tau_{3x}\\
\Delta \tau_{1y}-\Delta \tau_{3y}& \Delta \tau_{2y}-\Delta \tau_{3y}
\end{matrix}
\right).
\end{equation}
The matrix $\mathcal{A}$ changes to $\mathcal{A}'=\mathcal{A}+\Delta\mathcal{A}$, and we must modify $\mathcal{B}$ so that Eqn.~\eqref{eq:realreciprocal} still holds under mechanical load. To first order in displacements $\mathcal{A}'^{-1}$ becomes:
\begin{equation}\label{eq:correction}
\mathcal{A}'^{-1}=(1+\mathcal{A}\Delta\mathcal{A})^{-1}(\mathcal{A}^{-1})\simeq \mathcal{A}^{-1}-\mathcal{A}^{-1}\Delta\mathcal{A}\mathcal{A}^{-1}.
\end{equation}
By comparing Eqns. (7) and ~\eqref{eq:correction}, the reciprocal lattice vectors in Fig.~\ref{fig:F3}(b) must be renormalized by:
\begin{equation}
\Delta\mathcal{B}=-2\pi\left(\mathcal{A}^{-1}\Delta\mathcal{A}\mathcal{A}^{-1}\right)^T.
\end{equation}
 We note that the existence of this additional term is quite evident when working directly on the atomic lattice, but it was missed in Ref.~\cite{Kitt2012}, where the theory was expressed on a continuum. Let us now calculate some shifts of the $K-$points due to strain. For example, $\mathbf{K}_2$ ($=K$ in Fig.~\ref{fig:F3}(a)) requires an additional contribution, which we find by explicit calculation to be:
$$
\Delta K=\Delta\mathbf{K}_2=-\frac{4\pi}{3a_0^2}
\left(\Delta\tau_{1x}-\Delta\tau_{2x},\frac{\Delta \tau_{1x}+\Delta \tau_{2x}-2\Delta \tau_{3x}}{\sqrt{3}}\right),
$$
and using Eqn. (10) one immediately sees that $\Delta K'=-\Delta\mathbf{K}_2$, so that the $K$ ($\mathbf{K}_2$) and $K'$ ($-\mathbf{K}_2$) points shift in opposite directions, as expected \cite{GuineaNatPhys2010,castroRMP}.
\subsection{Gauge fields}

Equation \eqref{eq:applicabilitycondition} gives a condition for which the mechanical strain that varies smoothly on the scale of interatomic distances does not break the sublattice symmetry \cite{GuineaNatPhys2010}. On the other hand, arbitrary strain breaks down to some extent the periodicity of the lattice, and ``short-range'' strain can be identified to occur at unit cells where $\Delta \alpha_j$ and $\Delta L_j$ cease to be zero by significant margins.

 This observation provides the rationale for expressing the gauge fields without ever leaving the atomic lattice: When $\Delta \boldsymbol{\tau}_j'\simeq\Delta \boldsymbol{\tau}_j$ at each unit cell a mechanical distortion can be considered ``long-range,'' and the first-order theory is valid. The process to lay down the gauge terms to first order is straightforward. Local gauge fields can be computed as low energy approximations to the following $2\times 2$ pseudospin Hamiltonian:
\begin{equation}\label{eq:tbh}
\left(
\begin{matrix}
E_{s,A} & g^*\\
g & E_{s,B}
\end{matrix}
\right),
\end{equation}
with $g\equiv -\sum_{j=1}^3(t+\delta t_j)e^{i(\boldsymbol{\tau}_j+\Delta\boldsymbol{\tau}_j)\cdot(\mathbf{K}_n+\Delta\mathbf{K}_n+\mathbf{q})}$, and $n=1,...,6$. We defer discussion of the diagonal terms for now.

 Keeping exponents to first order we have:
$$
\small
(\boldsymbol{\tau}_j+\Delta\boldsymbol{\tau}_j)\cdot(\mathbf{K}_n+\Delta\mathbf{K}_n+\mathbf{q})\simeq
\boldsymbol{\tau}_j\cdot\mathbf{K}_n+\boldsymbol{\tau}_j\cdot\Delta\mathbf{K}_n+\Delta\boldsymbol{\tau}_j\cdot\mathbf{K}_n+
\boldsymbol{\tau}_j\cdot\mathbf{q}.
$$
The exponent is next expressed to first-order:
\begin{eqnarray}
e^{i(\boldsymbol{\tau}_j\cdot\mathbf{K}_n+\boldsymbol{\tau}_j\cdot\Delta\mathbf{K}_n+\Delta\boldsymbol{\tau}_j\cdot\mathbf{K}_n+
\boldsymbol{\tau}_j\cdot\mathbf{q})}\simeq \nonumber\\
ie^{i\boldsymbol{\tau}_j\cdot\mathbf{K}_n}\boldsymbol{\tau}_j\cdot\mathbf{q}+
e^{i\boldsymbol{\tau}_j\cdot\mathbf{K}_n}[1+i(\boldsymbol{\tau}_j\cdot\Delta\mathbf{K}_n+\Delta\boldsymbol{\tau}_j\cdot\mathbf{K}_n)].
\end{eqnarray}
Carrying out explicit calculations, one can see that:
\begin{equation}\label{eq:cancellation}
\sum_{j=1}^3e^{i\boldsymbol{\tau}_j\cdot\mathbf{K}_n}[1+i(\boldsymbol{\tau}_j\cdot\Delta\mathbf{K}_n+\Delta\boldsymbol{\tau}_j\cdot\mathbf{K}_n)]=0.
\end{equation}

For example, at $K=\mathbf{K}_2$ we have:
$$
\left[1+\frac{4i\pi(\Delta \tau_{1x}+\Delta \tau_{2x}+\Delta \tau_{3x})}{9a_0}\right](1+e^{\frac{2\pi i}{3}}-e^{\frac{\pi i}{3}}),
$$
with phasors adding up to zero. Similar phasor cancelations occur at every other $K-$point.

The term linear on $\Delta \mathbf{K}_n$ on Eqn.~\ref{eq:cancellation} cancels out the fictitious $K-$point dependent gauge fields proposed in Ref.~\cite{Kitt2012}, which originated from the term linear on $\Delta \mathbf{\tau}_j$ on this same equation. This observation constitutes yet another reason for the formulation of the theory directly on the atomic lattice. With this we have demonstrated that gauges will not depend explicitly on $K-$points, so we now continue formulating the theory considering the $\mathbf{K}_2$ point only \cite{GuineaNatPhys2010,Vozmediano,castroRMP}.

 Equation~\eqref{eq:tbh} takes the following form to first order at $\mathbf{K}_2$ in the low-energy regime:
\begin{eqnarray}\label{eq:ps1}
\mathcal{H}_{ps}=&
\left(
\begin{smallmatrix}
0 & t\sum_{j=1}^3ie^{-i\mathbf{K}_2\cdot\boldsymbol{\tau}_j}\boldsymbol{\tau}_j\cdot\mathbf{q}\\
-t\sum_{j=1}^3ie^{i\mathbf{K}_2\cdot\boldsymbol{\tau}_j}\boldsymbol{\tau}_j\cdot\mathbf{q} & 0
\end{smallmatrix}
\right)\nonumber\\
+&\left(
\begin{smallmatrix}
E_{s,A} & -\sum_{j=1}^3\delta t_je^{-i\mathbf{K}_2\cdot\boldsymbol{\tau}_j}\\
-\sum_{j=1}^3\delta t_je^{i\mathbf{K}_2\cdot\boldsymbol{\tau}_j} & E_{s,B}
\end{smallmatrix}
\right),
\end{eqnarray}
 with the first term on the right-hand side reducing to the standard pseudospin Hamiltonian in the absence of strain. The change of the hopping parameter $t$ is related to the variation of length, as explained in Refs.~\cite{Ando2002} and \cite{Vozmediano}:
\begin{equation}
\delta t_j=-\frac{|\beta| t}{a_0^2} \boldsymbol{\tau}_j\cdot\Delta\boldsymbol{\tau}_j.
\end{equation}
This way Eqn.~\eqref{eq:ps1} becomes:
\begin{eqnarray}
\mathcal{H}_{ps}=
\hbar v_F\boldsymbol{\sigma}\cdot \mathbf{q}
+\left(
\begin{smallmatrix}
E_{s,A} & f_1^*\\
f_1 & E_{s,B}
\end{smallmatrix}
\right),
\end{eqnarray}
with $f_1^*=\frac{|\beta|t}{2a_0^2}
[2\boldsymbol{\tau}_3\cdot\Delta\boldsymbol{\tau}_3
-\boldsymbol{\tau}_1\cdot\Delta\boldsymbol{\tau}_1
-\boldsymbol{\tau}_2\cdot\Delta\boldsymbol{\tau}_2
+\sqrt{3}i(\boldsymbol{\tau}_2\cdot\Delta\boldsymbol{\tau}_2-\boldsymbol{\tau}_1\cdot\Delta\boldsymbol{\tau}_1)]$, and $\hbar v_F\equiv
\frac{\sqrt{3}a_0t}{2}$.
The parameter $f_1$ can be expressed in terms of a vector potential: $A_s$ $f_1=-\hbar v_F\frac{eA_s}{\hbar}$. This way:
\begin{eqnarray}\label{eq:Asdiscrete}
\small
A_s&=-\frac{|\beta|\phi_0}{\pi a_0^3}[
\frac{2\boldsymbol{\tau}_3\cdot\Delta\boldsymbol{\tau}_3
-\boldsymbol{\tau}_1\cdot\Delta\boldsymbol{\tau}_1
-\boldsymbol{\tau}_2\cdot\Delta\boldsymbol{\tau}_2}{\sqrt{3}}\nonumber\\
&-i(
\boldsymbol{\tau}_2\cdot\Delta\boldsymbol{\tau}_2
-\boldsymbol{\tau}_1\cdot\Delta\boldsymbol{\tau}_1)].
\end{eqnarray}

We finally analyze the diagonal entries in Eqn.~\eqref{eq:tbh}, which are given as follows \cite{us}:
\begin{equation}\label{eq:EsA}
E_{s,A}=-\frac{0.3 eV}{0.12}\frac{1}{3}\sum_{j=1}^3\frac{|\boldsymbol{\tau}_j-\Delta\boldsymbol{\tau}_j|-a_0/\sqrt{3}}{a_0/\sqrt{3}},
\end{equation}
and
\begin{equation}\label{eq:EsB}
E_{s,B}=-\frac{0.3 eV}{0.12}\frac{1}{3}\sum_{j=1}^3\frac{|\boldsymbol{\tau}_j-\Delta\boldsymbol{\tau}'_j|-a_0/\sqrt{3}}{a_0/\sqrt{3}}.
\end{equation}
These entries represent the scalar deformation potential which we take to linear order in the average bond increase \cite{YWSon}.

\subsection{Relation to the formalism from first-order continuum elasticity}
 We next establish how the theory based on a continuum relates to the present formalism. In the absence of significant curvature, the continuum limit is achieved when $\frac{|\Delta\boldsymbol{\tau}_j|}{a_0}\to 0$ (for $j=1,2,3$). We have then (Cauchy-Born rule):
$\boldsymbol{\tau}_j\cdot \Delta \boldsymbol{\tau}_j\to \boldsymbol{\tau}_j\left(
\begin{smallmatrix}
u_{xx}&u_{xy}\\
u_{xy}&u_{yy}
\end{smallmatrix}\right)\boldsymbol{\tau}_j^T$, where $u_{ij}$ are the entries of the strain tensor.

 This way Eqn.~\eqref{eq:Asdiscrete} becomes:
\begin{equation}\label{eq:limit}
A_s\to \frac{|\beta|\phi_0}{2\sqrt{3}\pi a_0}(u_{xx}-u_{yy}-2iu_{xy}),
\end{equation}
as expected \cite{GuineaNatPhys2010,Vozmediano}.

Equation \eqref{eq:limit} confirms that if the zigzag direction is parallel to the $x-$axis the vector potential we have obtained is consistent with known results in the proper limit \cite{GuineaNatPhys2010,Vozmediano}.
 Besides representing a consistent first-order formalism, the present approach is exceptionally suited for the analysis of ``raw'' atomistic data --obtained, for example, from molecular dynamics simulations-- as there is no need to determine the strain tensor explicitly: the relevant equations (\ref{eq:Asdiscrete}, \ref{eq:EsA}, \ref{eq:EsB}) take as input the changes in atomic positions upon strain. Within the present approach $N/2$ space-modulated pseudospinor Hamiltonians can be built for a graphene membrane having $N$ atoms.

\section{Applying the formalism to rippled graphene membranes}

 We finish the present contribution by briefly illustrating the formalism on two experimentally relevant case examples. The developments presented here are motivated by recent experiments where freestanding graphene membranes are studied by local probes \cite{usold,stmNanoscale2012,stroscio}. (One must keep in mind, nevertheless, that the theory provided up to this point is rather general.)

\subsection{Rippled membranes with no external mechanical load}

 It is an established fact that graphene membranes will be naturally rippled due to a number of physical processes, including temperature-induced (i.e., dynamic) structural distortions \cite{Fasolino1}, and static structural distortions created by the mechanical and electrostatic interaction with a substrate, a deposition process \cite{Nature2007}, or line stress at the edges of finite-size membranes \cite{us}.

 In reference \cite{deJuanPRB} it is argued that the rippled texture of freestanding graphene leads to observable consequences, the strongest being a sizeable velocity renormalization. In order to demonstrate such statement, one must take a closer look at the underlying mechanics of the problem. The model \cite{deJuanPRB} assumes that a graphene membrane is originally pre-strained (in bringing an analogy, one would say that the membrane would be an ``ironed tablecloth''), so that curvature due to a single wrinkle directly leads to increases in interatomic distances. Those distance increases directly modify the metric on the curved space. In practice, an external electrostatic field can be used to realize such pre-strained configuration \cite{Fogler}.

\begin{figure}[tb]
\begin{center}
\includegraphics[width=0.48\textwidth]{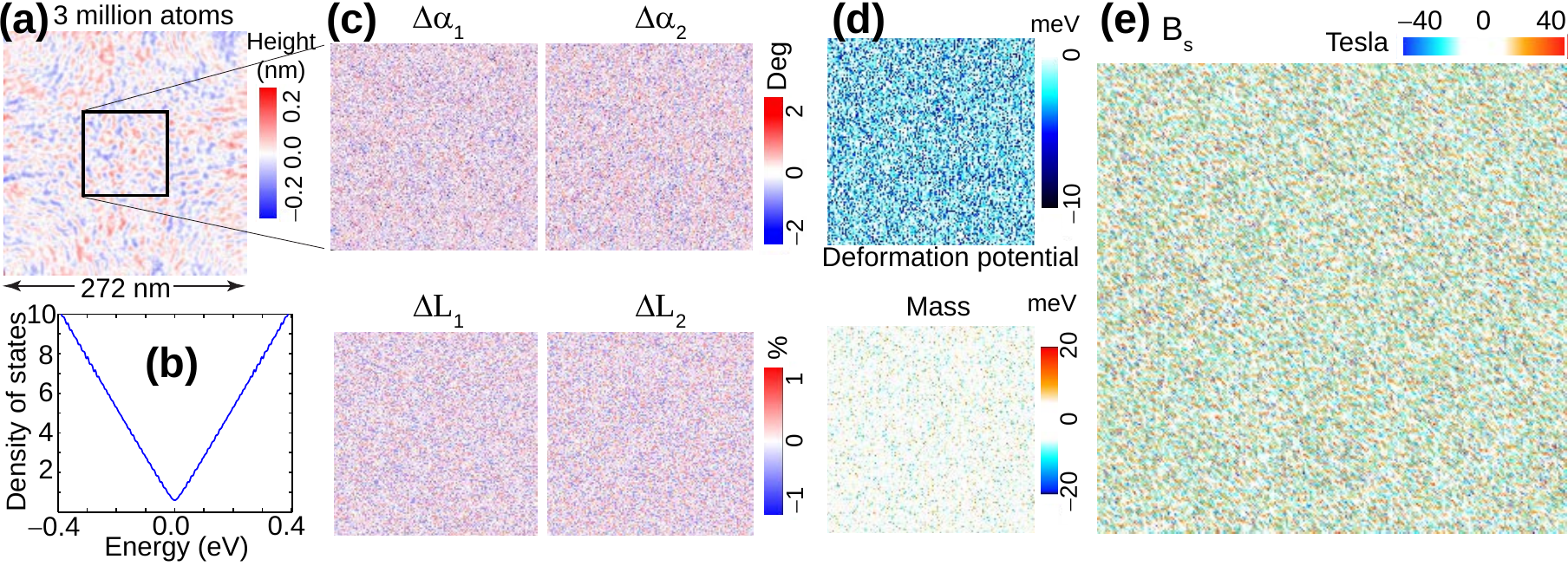}
\end{center}
\caption{A finite-size graphene membrane at 1 Kelvin. (a) The membrane forms ripples to relieve mechanical strain originating from its finite size. (b) We could not discern changes on the LDOS (which relates to renormalization of the Fermi velocity) on a completely flat membrane and after line strain is relieved. (c) Measures for changes in angles and lengths at individual unit cells (Eqns.~4-6) displaying noise on a small scale, and consistent with the formation of ripples. (d) The deformation potential, mass term and (e) the pseudo-magnetic field are inherently noisy as well.}\label{fig:F4}
\end{figure}

 In improving the consideration of the mechanics beyond first-order continuum elasticity, let us consider what happens if this pre-strained assumption is relaxed (in continuing our analogy, the rippled membrane in Fig.~\ref{fig:F4}(a) would then be akin to a ``wrinkled tablecloth prior to ironing''): How do the gauge fields look in such scenario? With our formalism, we can probe the interrelation between mechanics and the electronic structure directly. In Figure \ref{fig:F4}(a) we display a graphene membrane with three million atoms at 1 Kelvin after relaxing strain at the edges. The strain relaxation proceeds by the formation of ripples or wrinkles on the membrane. This initial configuration is already different to a flat (``pre-strained'') configuration within the continuum formalism, customarily enforced prior to the application of strain.

 \emph{The ripples must be ``ironed out'' before any significant increase on interatomic distances can occur:} ``Isometric deformations'' lead to curvature without any increase on interatomic distances \cite{us} (in continuing our analogy, this is usually what happens with clothing). We believe that a local determination of the metric tensor from atomic displacements alone will definitely be useful in continuing making a case for velocity renormalization \cite{deJuanPRL2012,arxiv,deJuanPRB}; this is presently work in progress \cite{us2}.

 The local density of electronic states is obtained directly from the Hamiltonian of the membrane in configuration space $\mathcal{H}$, and shown in Fig.~\ref{fig:F4}(b). When compared to the DOS from a completely flat membrane, no observable variation on the slope of the DOS appears, and hence, no renormalization of the Fermi velocity either.

One can determine the extent to which nearest-neighbor vectors will preserve sublattice symmetry in terms of $\Delta\alpha_j$ and $\Delta L_j$, Eqns.~(4-6). We observe small and apparently random fluctuations on those measures in Fig.~\ref{fig:F4}(c): $\Delta L_j\lesssim  $ 1\% and $\Delta \alpha_j\lesssim   2^{o}$.

We display the deformation potential in Figure \ref{fig:F4}(d) in terms of the average ($E_{def}$) and difference ($E_{mass}$)
between $E_{s,A}$ and $E_{s,B}$ (Eqns.~(\ref{eq:EsA}) and (\ref{eq:EsB})) at any given unit cell:
\begin{equation}
E_{def}=\frac{1}{2}(E_{s,A}+E_{s,B}), \text{ and } E_{mass}=\frac{1}{2}(E_{s,A}-E_{s,B}).
\end{equation}
Both quantities are of the order of tens of meVs.

The ripples lead to the random-looking pseudo-magnetic field shown in Fig.~\ref{fig:F4}(e), reminiscent of the electron density plots created by random charge puddles \cite{Rossi1,Rossi2}. 
 We next consider how strain by a sharp probe modifies the results in Fig.~\ref{fig:F4}.

\subsection{Rippled membranes under mechanical load}
\begin{figure}[tb]
\begin{center}
\includegraphics[width=0.49\textwidth]{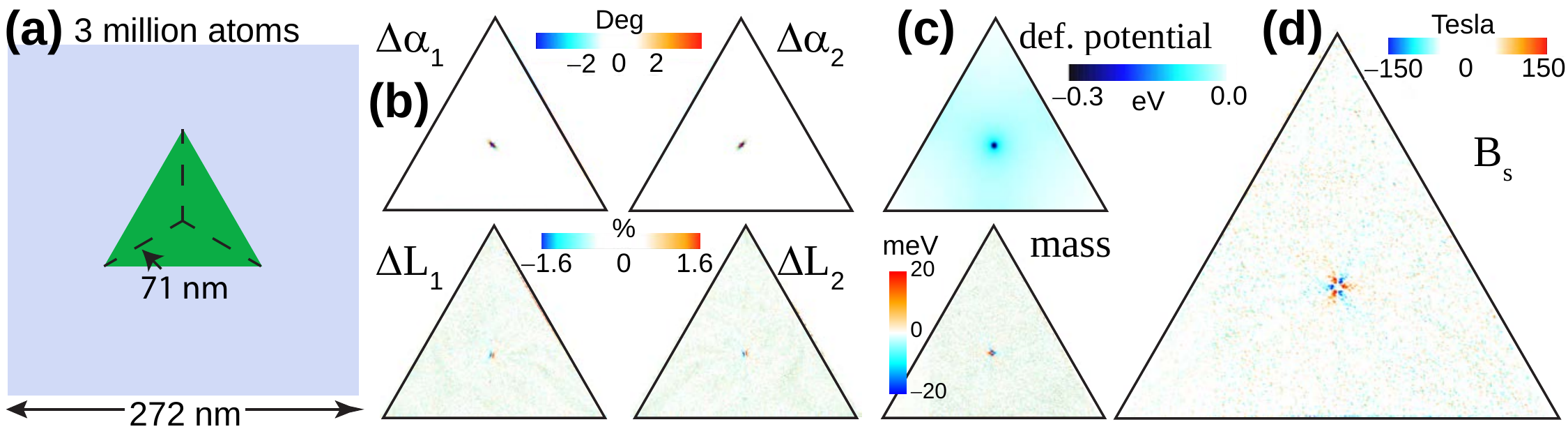}
\end{center}
\caption{Strained membrane: (a) The section in blue (light gray) is kept fixed, and strain is applied by pushing down the triangular section in green (dark gray) with a sharp extruder, located at the geometrical center. (b) Deviations from proper sublattice symmetry are concentrated at the section directly underneath the sharp tip, where the deformation is the largest and strain is the most inhomogeneous. (c-d) Gauge fields.}\label{fig:F5}
\end{figure}

In what follows we consider a central extruder creating strain on the freestanding membrane. For this, we placed the membrane shown in Fig.~\ref{fig:F4} on top of a substrate (shown in blue/light gray in Fig.~\ref{fig:F5}(a)) with a triangular-shaped hole (in green/dark gray in Fig.~\ref{fig:F5}(a)). The membrane is held fixed in position when on the substrate, and pushed down by a sharp tip at its geometrical center, down to a distance $\Gamma$=10 nm.
As indicated earlier, sublattice symmetry is not exactly satisfied right underneath the tip, where $\Delta\alpha_j$ and $\Delta L_j$ take their largest values (Fig.~\ref{fig:F5}(c)). While $\Delta L_j$ still displays some fluctuations, this is not the case for $\Delta \alpha_j$ (the scale for $\Delta \alpha_j$ is identical to that from Fig.~\ref{fig:F4}(c)). The large white areas tells us that fluctuations on $\Delta\alpha_j$ are wiped out upon load as the extruder removes wrinkles. This observation stems from the lattice-explicit consideration of the mechanics.

We have presented a detailed discussion of the problem along these lines \cite{us}. We found that for small magnitudes of load a rippled membrane will adapt to an extruding tip isometrically. This observation is important in the context of the formulation with curvature \cite{deJuanPRB,deJuanPRL2012}, because in that formulation there is the assumption that distances between atoms increase as soon as graphene deviates from a perfect 2-dimensional plate.

 The gauge fields given in Fig.~\ref{fig:F5}(c-d) reflect the circular symmetry induced by the circular shape of the extruding tip~\cite{us}.

\begin{figure}[h!]
\begin{center}
\includegraphics[width=0.48\textwidth]{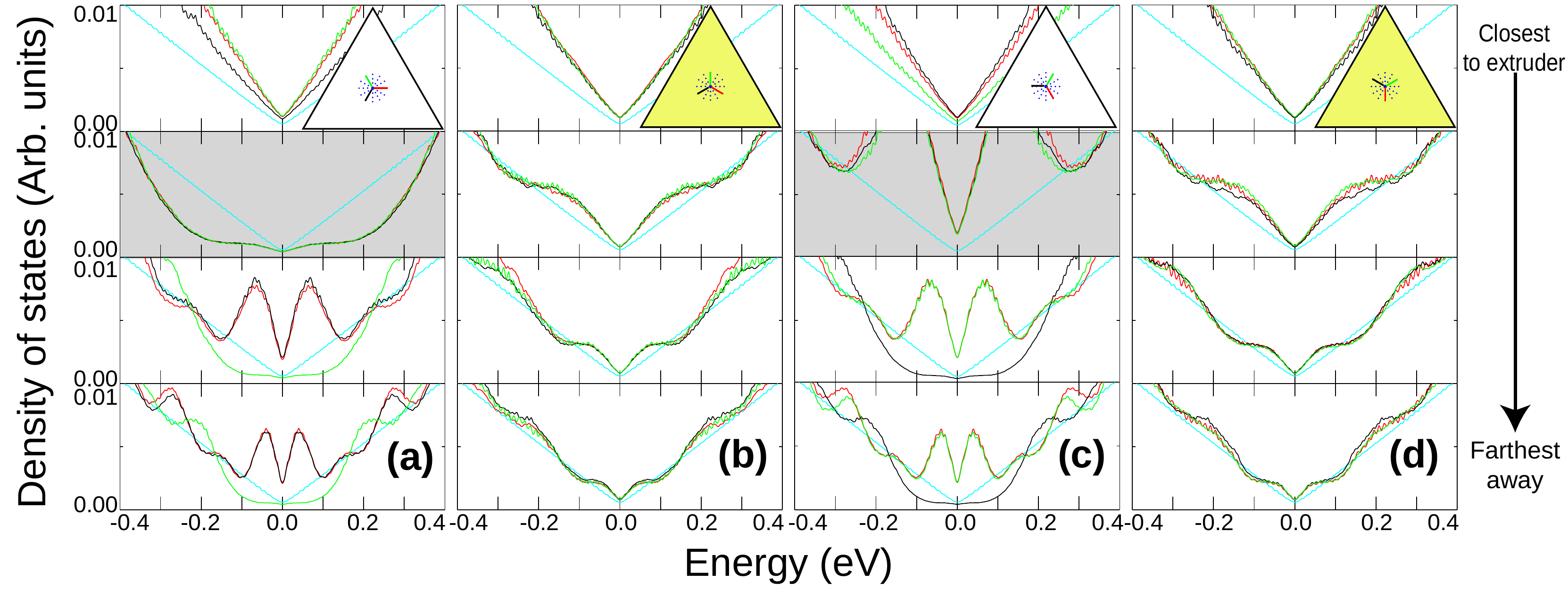}
\end{center}
\caption{Local density of states on the membrane under strain shown in Fig.~\ref{fig:F5}. The locations where the DOS
is computed are shown in the insets (the most symmetric line patterns are displayed in yellow).}\label{fig:F6}
\end{figure}
We finish the discussion by probing the local density of states at many locations in Fig.~\ref{fig:F6}, which may relate to the discussion of confinement by gauge fields \cite{Blanter}. $E_s$ was was not included in computing DOS curves.

Some generic features of DOS are clearly visible: (i) Near the extruder, the deformation is already beyond the linear regime, and the DOS is indeed renormalized for locations close to the mechanical extruder \cite{deJuanPRL2012,arxiv,deJuanPRB}. (ii) A sequence of features appear on the DOS farther away from the extruder. Because the field is not homogeneous and perhaps due to energy broadening we are unable to tell a central peak. As indicated on the insets, the plots on Fig.~6(b) and 6(d) are obtained along high-symmetry lines (the colors on the DOS subplots correspond with the colored lines on the insets). For this reason they look almost identical, and the three sets of curves (corresponding to the DOS along different lines) overlap. Due to lower symmetry, the LDOS in Fig.~6(a) and 6(c) appear symmetric in pairs, with the exception of the plots highlighted in gray. (the light 'v'-shaped curve in all subplots is the reference DOS in the absence of strain).

 LDOS curves complement the insight obtained from gauge field plots. Hence, they should also be reported in discussing strain engineering of graphene's electronic structure, particularly in situations where gauge fields are inhomogeneous.

\section{Conclusions}
We presented a novel framework to study the relation between mechanical strain and the electronic structure of graphene membranes. Gauge fields are expressed directly in terms of changes in atomic positions upon strain. Within this approach, it is possible to determine the extent to which the sublattice symmetry is preserved. In addition, we find that there are no $K-$dependent gauge fields in the first-order theory. We have illustrated the method by computing the strain-induced gauge fields on a rippled graphene membrane with and without mechanical load. In doing so, we have initiated a necessary discussion of mechanical effects falling beyond a description within first-order continuum elasticity. Such analysis is relevant for accurate  determination of gauge fields and has not received proper attention yet.\\

\noindent{\bf Acknowledgments}\\
We acknowledge conversations with B. Uchoa, and computer support from HPC at Arkansas (\emph{RazorII}), and XSEDE (TG-PHY090002, \emph{Blacklight}, and \emph{Stampede}). M.V. acknowledges support by the Serbian Ministry of Science, Project No. 171027.

\end{document}